\newif\ifhires \hiresfalse

\documentclass[epj-spec]{svjour}
\usepackage{graphicx,float}
\usepackage{amsmath}

\newcommand{\w}{\omega}

\newcommand{\Tc}{T_{\rm c}}

\newcommand{\Qch}{{\vec Q}_{\rm c}}
\newcommand{\Qchx}{{\vec Q}_{{\rm c}x}}
\newcommand{\Qchy}{{\vec Q}_{{\rm c}y}}
\newcommand{\Qsp}{{\vec Q}_{\rm s}}
\newcommand{\Qspx}{{\vec Q}_{{\rm s}x}}
\newcommand{\Qspy}{{\vec Q}_{{\rm s}y}}

\newcommand{\ybco}{YBa$_2$Cu$_3$O$_{6+\delta}$}

\newcommand{\lbco}{La$_{2-x}$Ba$_x$CuO$_4$}
\newcommand{\lbcoo}{La$_{15/8}$Ba$_{1/8}$CuO$_4$}
\newcommand{\lsco}{La$_{2-x}$Sr$_x$CuO$_4$}

\newcommand{\bscco}{Bi$_2$Sr$_2$CaCu$_2$O$_{8+\delta}$}
\newcommand{\ccoc}{Ca$_{2-x}$Na$_x$CuO$_2$Cl$_2$}

\begin{document}
\title{
Tendencies toward nematic order in \ybco:\\
Uniform distortion vs. incipient charge stripes}
\author{Matthias Vojta}
\institute{
Institut f\"ur Theoretische Physik, Universit\"at zu K\"oln,
Z\"ulpicher Stra\ss e 77, 50937 K\"oln, Germany
}

\abstract{
Recent neutron scattering and transport data obtained on underdoped \ybco,
with strong signatures of rotation symmetry breaking at low temperatures, point toward
electron-nematic order in the charge sector. Such order may originate from a uniform
distortion with $d$-wave symmetry or as a precursor of a uni-directional stripe phase.
Here, we discuss whether the neutron scattering data can be linked to incipient charge
stripes. We employ and extend a phenomenological model for collective spin and charge
fluctuations and analyze the resulting spin excitation spectrum under the influence of
lattice anisotropies. Our results show that the experimentally observed
temperature-dependent magnetic incommensurability is compatible with a scenario of
incipient stripes, the temperature dependence being due to the temperature variation of
both strength and correlation length of the charge stripes. Finally, we propose further
experiments to distinguish the possible theoretical scenarios.
}

\maketitle


\section{Introduction}
\label{intro}

The interplay of different ordering tendencies is a recurring theme in unconventional
superconductors -- this applies to cuprates, iron pnictides, and heavy-fermion
superconductors. In high-$\Tc$ superconducting cuprates, phenomena of spontaneous lattice
symmetry breaking have been observed inside the pseudogap regime
\cite{stripe_rev1,stripe_rev2}. In particular, stripe order in both the spin and charge
sectors has been established in the \lsco\ (or 214) family \cite{jt95}, while in \ybco\
(YBCO), signatures of electronic nematic order have been reported
\cite{hinkov08a,taill10b}. Given the similarities in the spin excitation spectra at
intermediate energies -- the so-called hour-glass spectrum -- across different cuprate
families, it has long been proposed that stripes might be a common origin of
incommensurate low-energy spin fluctuations \cite{stripe_rev1,stripe_rev2,jt05}. This
view has been fueled by the observation of stripe-like modulations in the charge sector
on the surface of \bscco\ and \ccoc\ using scanning tunneling microscopy (STM)
\cite{kapi03a,yazdani04,kohsaka07}. It has been argued that a picture of spatially disordered (i.e. slowly
fluctuating or disorder-pinned) stripes is broadly consistent with both neutron and STM
data \cite{vvk,mv08}. In this scenario, modulations in the charge sector are driving
incommensurabilities in the spin sector. This view is compatible with the phase diagrams
of various 214 cuprates, where charge order is established at a higher temperature than spin
order, i.e., charge stripes can exist without ordered magnetism over a range of
temperatures.

Recent experimental data on de-twinned YBCO, where both neutron scattering \cite{hinkov08a} and
thermoelectric transport \cite{taill10b} data show distinct in-plane anisotropies below a
doping-dependent onset temperature, have suggested to consider electron-nematic order,
i.e., rotation symmetry breaking with translation symmetry preserved, as a separate
player. In fact, simple model calculations for the anisotropy of the spin excitations
\cite{lawler10} and that of the Nernst signal \cite{hv09} based on uniform distortions
describing nematic order were found to describe salient aspects of the data.\footnote{
Note that the YBCO crystal structure is orthorhombic due
to the presence of CuO chains, such that the in-plane rotation symmetry is broken from
the outset. Hence, there is no symmetry-breaking electronic phase transition, but the common
assumption is that the structural anisotropies are enhanced by correlation
effects at low temperatures, signaling the tendency toward electron-nematic order.
}
This naturally prompts the question whether nematic or stripe order should be considered
as the ``primary'' order parameter. This issue is of importance, as it pertains to
proposals which link the origin of the pseudogap in underdoped cuprates to a
symmetry-breaking order competing with superconductivity.

In this paper, we shall attempt to answer the question whether the neutron data of
Ref.~\cite{hinkov08a} are consistent with a scenario of spatially disordered charge
stripes as advocated in Ref.~\cite{vvk}. To this end, we study the coupled
order-parameter field theory of Ref.~\cite{vvk} in a regime of a small spin gap under the
influence of lattice anisotropies. Using plausible assumptions for the temperature
dependence of the collective charge degrees of freedom, we determine the anisotropy of
the low-energy spin fluctuations and compare them to the data of Ref.~\cite{hinkov08a}.
Doing so, we find the answer to the above question to be a tentative ``yes''; hence, we
provide here a mechanism for the observed temperature-dependent incommensurability
\cite{hinkov08a} alternative to the one of Ref.~\cite{lawler10}. Finally, we propose
further experiments to discriminate between available theoretical scenarios.

The body of the paper is organized as follows: To set the stage, we shall begin with
general remarks about order parameters, symmetries, and phase diagrams in
Sec.~\ref{sec:symm}. We then turn to the spin dynamics and describe in
Sec.~\ref{sec:fluct} the coupled order-parameter theory of Ref.~\cite{vvk}, with
appropriate modifications. Sec.~\ref{sec:res} presents the numerical results for the
dynamic spin susceptibility, together with the analysis of the low-energy
incommensurability. A discussion of experimentally relevant issues and an outlook
will close the paper.


\section{Stripe and nematic order}
\label{sec:symm}

In the context of uni-directional stripe order in quasi-2d systems, order parameters for
spin and charge density waves as well as for rotational symmetry breaking can be
considered. In the following we shall discuss a continuum description which is
appropriate for slowly varying order-parameter fields. For reasons explained below, the
calculation in Sec.~\ref{sec:fluct} will instead use a lattice formulation.

The charge density wave (CDW) requires a pair of complex scalar fields $\phi_{cx}$,
$\phi_{cy}$ for the two CDW directions with wavevector $\Qchx$ and $\Qchy$.\footnote{
This continuum description cannot capture modulations with wavevectors significantly
different from $\Qchx$ and $\Qchy$.}
Then, the charge density follows
\begin{equation}
\langle \rho ({\vec R}, \tau) \rangle = \rho_{\rm avg} + \mbox{Re} \left[e^{i \Qch
\cdot {\vec R}} \phi_c (\vec R, \tau)  \right] \,.
\label{chargemod}
\end{equation}
Similarly, there is a pair of complex vector fields $\phi_{s\alpha x}$, $\phi_{s\alpha
y}$ describing spin density waves (SDW) with wavevectors $\Qspx$ and $\Qspy$. For
cuprates at dopings above 5\%, order has been found at $\Qspx = 2\pi(0.5\pm 1/M,0.5)$,
$\Qspy = 2\pi(0.5,0.5\pm 1/M)$ and $\Qchx = (2\pi/N,0)$, $\Qchy = (0,2\pi/N)$, where $M$
and $N$ are the real-space periodicities which follow $M=2N$ to a good accuracy
\cite{stripe_rev1,stripe_rev2,jt95}.
Finally, there is an Ising scalar $\phi_n$ for $l=2$ spin-symmetric electron-nematic
order at wavevector $\vec Q=0$. The dynamics of any of these order parameters should be
described by an appropriate $\phi^4$ (or Landau-Ginzburg-Wilson) action. Other order
phenomena, e.g., uniform and modulated superconducting pairing, shall not be of interest
here.

Biquadratic local couplings between all order parameters are generically symmetry
allowed. For instance, a term $v|\phi_{cx}(\vec r,\tau)|^2|\phi_{cy}(\vec r,\tau)|^2$,
which couples horizontal and vertical charge stripes, decides about repulsion or
attraction between $\phi_{cx,y}$, i.e., $v>0$ will lead to uni-directional (stripe) order
where $v<0$ results in bi-directional (checkerboard) order.

More interesting are couplings involving one order parameter {\em linearly}.
Those terms are strongly constrained by symmetry and momentum conservation.
A nematic order parameter $\phi_n$ in a tetragonal environment
couples to CDW and SDW according to
\begin{equation}
\kappa_1 \phi_n (|\phi_{cx}|^2-|\phi_{cy}|^2) +
\kappa_2 \phi_n (|\phi_{s\alpha x}|^2-|\phi_{s\alpha y}|^2),
\label{coupl1}
\end{equation}
note that $|\phi_{cx}|^2$ etc. carry vanishing lattice momentum.
A CDW order parameter couples to a spin density wave $\phi_s$ according to
\begin{equation}
\kappa_3 (\phi_c^\ast \phi_{s\alpha}^2 +c.c.)
\label{coupl2}
\end{equation}
which is allowed only if the ordering wavevectors obey $\Qch = 2\Qsp$.
In addition, symmetry-allowed higher-order couplings between gradients of order parameters
will occur.

\begin{figure}
\begin{center}
\includegraphics[width=5in]{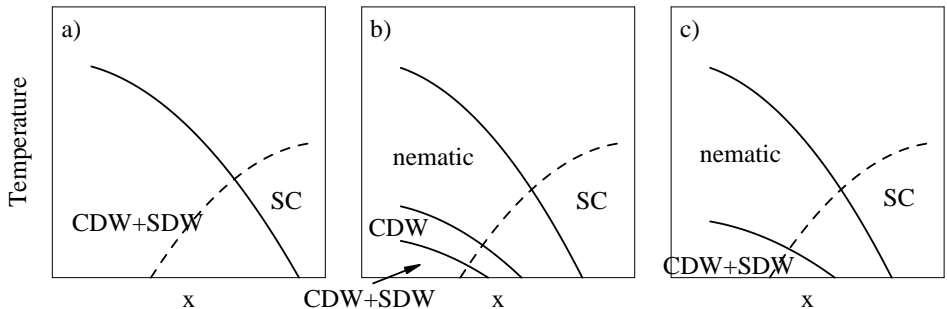}
\caption{
Schematic phase diagrams, illustrating possible transition scenarios into symmetry-broken
states. The vertical axis is temperature, whereas the horizontal axis -- where $x$ may represent doping --
tunes the interplay between superconductivity and spin/charge/nematic order, which are assumed to be
competitive. The solid (dashed) lines are transitions into lattice-symmetry-breaking (superconducting)
states.
Case a) corresponds to a weak-coupling scenario, with a single stripe ordering
transition, while b) may be realized at strong coupling, with distinct transitions for nematic,
charge, and spin order \cite{KFE98}.
Case c) has a nematic, but no CDW phase: in this scenario, the nematic order may be
stripe-free and eventually trigger a transition to an incommensurate SDW phase.
In all cases, quenched disorder will smear out both CDW and nematic transitions and likely
turn the SDW into a cluster spin glass \cite{ek93}.
Further, a structural anisotropy will smear out the nematic transition -- this applies
to the LTT phase of 214 cuprates and to YBCO.
Qualitatively, case b) might be realized in some 214 cuprates (with the nematic
transition replaced by the LTO-LTT transition), whereas cases b) or c) might apply to
YBCO.
}
\label{fig:pd2}
\end{center}
\end{figure}

The terms (\ref{coupl1},\ref{coupl2}) imply that a uni-directional density wave induces
nematic order via $\kappa_{1,2}$ and that a collinear SDW with $\Qsp$ induces a CDW with
$2\Qsp$ via $\kappa_3$. As a consequence, a transition into a stripe-ordered state may
occur as a direct transition from a disordered into a CDW+SDW state, or via intermediate
nematic and CDW phases, see Fig.~\ref{fig:pd2}. The phase diagram of the corresponding
Landau theory has been worked out in detail by Zachar {\em et al.} \cite{zachar98}.
Whether the intermediate phases are realized depends on microscopic details.
In particular, an electron-nematic phase can come in (at least) two flavors, namely (i) a
purely uniform ($\vec Q=0$) distortion, as arises from a Pomeranchuk instability of the
Fermi surface (dubbed {\em ``stripe-free''} nematic in the following), or (ii) a
precursor to a uni-directional stripe phase, in which case it will display substantial
and slow charge fluctuations at non-zero wavevectors (hence dubbed {\em
``stripe-driven''}). Although stripe-free and stripe-driven nematics do not differ in
symmetries, the underlying microscopic mechanisms are clearly distinct.

While weak-coupling theories typically give a direct transition into a CDW+SDW state,
Fig.~\ref{fig:pd2}a, distinct scenarios are viable at strong coupling \cite{KFE98}. For instance, the
tendency toward bond order could be dominant \cite{vs}, which induces both translation
and rotation symmetry breaking in the charge sector. Incommensurabilities in the spin
sector would then follow from periodic modulations of the charge, as in
Fig.~\ref{fig:pd2}b. Alternatively, stripe-free electron-nematic order might be the
primary phenomenon, which then drives ordering phenomena in both the charge and spin
sectors, Fig.~\ref{fig:pd2}c.

For 214 cuprates with LTT lattice structure, the in-plane rotation symmetry is broken in
each CuO$_2$ layer, such that there can be no transition into a purely nematic phase.
Experimentally, distinct transition are seen in the charge and spin sectors in \lbco\ as
well as in \lsco\ co-doped with Nd or Eu. These transitions can be associated with the
onset of CDW and SDW order, respectively. Remarkably, these transitions apparently
continue to exist in \lbcoo\ under high pressure where the material remains in the
so-called HTT phase, without structurally broken in-plane rotation symmetry, down to
lowest temperatures \cite{huecker10}. In contrast, in YBCO, no signatures of (static)
charge order have been reported. Static spin order is restricted to underdoped samples
and very low temperatures: in YBCO-6.45, quasi-static incommensurate order in neutron
scattering sets in below 30\,K, with a correlation length of about 20\,\AA\ only, and
$\mu$SR measurements detect static moments only at 1.5\,K, suggesting that the order is
weak and glassy \cite{hinkov08a}. However, signatures of rotation symmetry breaking are
prominent in both the spin excitations \cite{hinkov08a} and in the Nernst effect
\cite{taill10b}. Although YBCO also displays a structural anisotropy, smearing out a
transition into an electron-nematic state, the experimentally observed anisotropies
display a relatively well-defined doping-dependent onset temperature which has been
proposed to follow the pseudogap temperature.

At present, it is unclear whether the nematic order in YBCO should be viewed as
stripe-free or stripe-driven.
In the literature, different points of view have been put forward. On the one hand,
simple stripe-free models of electron-nematic order have been invoked to explain the
anisotropic incommensurate spin excitation spectrum in YBCO as detected in
Refs.~\cite{hinkov08a,hinkov07}. Here, both RPA on top of an anisotropic band structure
\cite{eremin05a,schnyder06,yamase06,yamase08} and a Ginzburg-Landau theory for spin
excitations with phenomenological anisotropy \cite{lawler10} have been employed. A model
of electrons moving in an anisotropic band structure has been found \cite{hv09} to
describe some features of the Nernst effect measured in field up to 15\,T in YBCO
\cite{taill10b}. The Nernst analysis also suggests that stripe order sets in at lower
temperatures \cite{hvs10,taill09}.
On the other hand, the notion of apparently ``universal'' spin excitations at
intermediate energies in various cuprates \cite{jt05} has triggered stripe-based
explanations. In particular, fluctuating or disorder-pinned stripes have been shown to
induce a hour-glass-like spin excitation spectrum which is compatible with most of the
YBCO neutron scattering data. In this scenario, anisotropies in $\chi''$ arise
due to the small (chain-induced) bare electronic anisotropic which is strongly enhanced
due to the proximity to a uni-directionally ordered state.


\section{Lattice order parameter theory for disordered stripes}
\label{sec:fluct}

The remainder of the paper is concerned with the question of whether a scenario of disordered
stripes can be consistent with the neutron data obtained on YBCO-6.45 \cite{hinkov08a}.
In this section, we summarize the coupled order-parameter theory of Refs.~\cite{vvk,mvss06} and
discuss the parameter regime of potential relevance for the experiment. Numerical results
follow in the next section.

The approach of Refs.~\cite{vvk,mvss06} builds on a few main assumptions: (i) The collective
physics in both the spin and charge sectors can be captured by order-parameter fields,
without reference to single-particle excitations. (ii) The ``bare'' spin sector is
dominated by fluctuations near wavevector $(\pi,\pi)$, and any incommensurability in the
spin sector is therefore driven by its coupling to the charge sector.
(iii) In the presence of incommensurabilities, the spin order remains collinear.
Item (ii) implies
that neither a continuum formulation nor a perturbative treatment of the spin--charge
coupling are justified, because the spin ordering wavevector can change significantly
(and in a non-perturbative fashion) due to a coupling to the charge modes.
Items (i) and (ii) are borrowed from the successful description of magnetic excitations of
static stripes, as measured by Tranquada {\em et al.} \cite{jt04} in stripe-ordered
\lbcoo, using models of coupled spin ladders \cite{mvtu04,gsu04,seibold05,carlson06}.
Such models assume static charge order, which in turn imposes a strong modulation on both spin
density and magnetic couplings; without this modulation the spin sector is described by a
nearest-neighbor Heisenberg model.

For the lattice order-parameter description, we start from a $\varphi^4$ theory for the spin sector:
\begin{eqnarray}
\mathcal{S}_\varphi =
\int \! d \tau \sum_j \left[ (
\partial_\tau \vec\varphi_j)^2 +
s \vec\varphi_j^2 \right]
+  \sum_{\langle j j' \rangle} \! c^2
(\vec\varphi_j \!-\! \vec\varphi_{j'})^2 + \mathcal{S}_4
\label{sphi}
\end{eqnarray}
with $\mathcal{S}_4$ being the quartic self-interaction term.
The real field $\varphi_j \equiv \varphi(\vec{r}_j)$ is defined on the sites $j$ of the square
lattice of Cu atoms. It carries lattice momentum $\vec Q=(\pi,\pi)$, such that
$\varphi_j$ and the Cu spins $\vec S_j$ are related through
$
\vec S_j \propto e^{i {\vec Q} \cdot {\vec r}_j} \vec\varphi_j
$, and
$\varphi_j={\rm const.}$ describes the familiar two-sublattice antiferromagnet.
In the absence of static order, the quartic term $\mathcal{S}_4$ may be neglected, which
amounts to a Gaussian approximation to the spin fluctuations.

For the charge sector we introduce complex order parameters $\psi_{x,y}$
for stripe modulations with wavevectors $\Qchx$ and $\Qchy$ -- below the focus will be on
period-4 stripes. \footnote{
Linking the notations of Sec.~\ref{sec:symm} and
Ref.~\cite{vvk}, $\phi_c \equiv \psi$ and $\phi_s \equiv \varphi$.
However, $\psi$ and $\varphi$ are lattice instead of continuum fields.
}
Then, the real field
$
Q_x ({\vec r} ) = {\rm Re} \psi_x ({\vec r}) e^{i \Qchx \cdot {\vec r}}
$
(similarly for $Q_y$) measures the modulation of site and bond charge densities.
We choose signs such that
$\delta\rho({\vec r}_j) = Q_x + Q_y$ is the deviation of the local {\em hole} density
from its spatial average.
The allowed couplings between spin and charge are dictated by spin and time reversal
symmetries, with the lowest order ones being of the form  $\lambda Q
\vec\varphi^{\,2}$.
Guided by the lattice models \cite{mvtu04,gsu04} we choose \cite{mvss06}
\begin{eqnarray}
\mathcal{S}_{\varphi\psi} &=& \int d \tau \sum_j \Bigl[
\lambda_1 Q_x ({\vec r}_j ) \vec\varphi_{j}^2 + \lambda_2 Q_x
({\vec r}_{j+x/2})
\vec\varphi_{j} \vec\varphi_{j+x} \nonumber\\
&+& \lambda_3 Q_x ({\vec r}_j)
\vec\varphi_{j-x} \vec\varphi_{j+x} + \lambda_4 Q_x ({\vec
r}_{j+y/2}) \vec\varphi_{j} \vec\varphi_{j+y} \Bigr] \nonumber \\
&+&  \Bigl[ x \leftrightarrow y \Bigr].
\label{sphipsi}
\end{eqnarray}
Here, $\lambda_1>0$ implements the correlation between the on-site charge and spin
densities, i.e., a large hole density reduces the local amplitude of the spin
fluctuations. $\lambda_{2-4}$ ensure that the effective first- and second-neighbor
exchange constants modulate along with the charge/bond order. In particular, the
antiphase domain wall properties of the stripes \cite{pnas,doug} are reflected in the
positive sign of $\lambda_{2,3}$, e.g., for large $\lambda_2 Q_x$ the nearest-neighbor
magnetic coupling perpendicular to the stripe direction in $\mathcal{S}_\varphi +
\mathcal{S}_{\varphi\psi}$ becomes ferromagnetic (instead of antiferromagnetic).

Perfect static vertical (horizontal) stripe order corresponds to $\psi_x=\rm const$,
$\psi_y=0$ ($\psi_y=\rm const$, $\psi_x=0$). Then, the action $\mathcal{S}_\varphi +
\mathcal{S}_{\varphi\psi}$ is a theory for magnetic modes in a background of static
charge order. For sufficiently large $\lambda$ couplings, the minimum energy of the
$\varphi$ fluctuations will be shifted away from $(\pi,\pi)$ to the incommensurate
wavevector $\Qsp$ dictated by the charge order (which fulfills $2\Qsp = \Qch)$. A
Gaussian approximation to such a theory \cite{mvss06} reproduces the results obtained
from microscopic models of coupled two-leg ladders \cite{mvtu04,gsu04} which are in agreement with
experimental data on \lbcoo\ \cite{jt04}.

Deviations from perfect charge order can be described by a
$\psi^4$-type theory of the form
\begin{eqnarray}
\mathcal{S}_{\psi} &=& \int \!d\tau \! \sum_i \Bigl[
\left| \partial_\tau \psi_{ix} \right|^2 +
\left| \partial_\tau \psi_{iy} \right|^2 +
s_x |\psi_{ix}|^2 + s_y |\psi_{iy}|^2 \nonumber\\
&+&
 c_{1x}^2 | \psi_{ix}-\psi_{i+x,x} |^2 +
 c_{2x}^2 | \psi_{ix}-\psi_{i+y,x} |^2 +
 c_{1y}^2 | \psi_{iy}-\psi_{i+x,y} |^2 +
 c_{2y}^2 | \psi_{iy}-\psi_{i+x,y} |^2
\nonumber \\
&+&
u_1 \psi_i^4 + u_2 \psi_i^6
+ v |\psi_{ix}|^2 |\psi_{iy}|^2
+ w \left( \psi_{ix}^4 \!+\! \psi_{ix}^{\ast 4}
\!+\! \psi_{iy}^4 \!+\! \psi_{iy}^{\ast 4} \right) \Bigr]
\label{spsi}
\end{eqnarray}
with $\psi_{ix}\equiv \psi_x(\vec{r}_i)$ and $\psi_i^2\!\equiv\!|\psi_{ix}|^2\!+\!|\psi_{iy}|^2$.
This theory features a variety of non-linear terms, which have been included to model the
non-trivial physics of the underlying strongly-correlated system on the lattice scale
(recall that the objective is not to describe universal critical phenomena).
Here, a combination of $u_1\!<\!0$ and $u_2\!>\!0$ suppresses amplitude fluctuations
of $\psi$ -- this is motivated by the STM data of Ref.~\cite{kohsaka07}.
For $c_{1x}\!=\!c_{1y}$, $c_{2x}\!=\!c_{2y}$, $s_x\!=\!s_y$, and $v\!=\!w\!=\!0$,
the action has O(4) symmetry, which, however, is broken by various lattice effects:
The quartic $v |\psi_x|^2 |\psi_y|^2$ term regulates the repulsion or
attraction between horizontal and vertical stripes;
we employ $v\!>\!0$ leading to stripe-like order
(whereas $v\!<\!0$ results in checkerboard structures).
The phase-sensitive $w$ term provides commensurate pinning and
selects bond-centered (instead of site-centered)
stripes \cite{kohsaka07,vvk} for $w\!>\!0$.

The treatment of $\mathcal{S}_{\varphi}+\mathcal{S}_{\psi}+\mathcal{S}_{\varphi\psi}$
(\ref{sphi}--\ref{spsi})
requires further approximations. Ref.~\cite{vvk} proposed an adiabatic approximation,
based on the assumption that the charge modes are slow compared to the spin modes. As in
the familiar Born-Oppenheimer approximation, the spin sector may then be treated for a
fixed configuration of charges, and a proper average over the charge configurations has
to be taken. For simplicity, one may further neglect the feedback of the spin sector on
the charge sector.
Practically, this then amounts to the following procedure: A classical Monte-Carlo
simulation of the action $\mathcal{S}_{\psi}$ (without the $\partial_\tau$ terms, at a
temperature $T=1$) is used to generate individual configurations of the charge order
parameter field $\psi_{x,y}$ (on a finite lattice of up to $64^2$ sites), from which one
calculates the $Q_{x,y}$. For each configuration of $Q_{x,y}$, $\mathcal{S}_\varphi +
\mathcal{S}_{\varphi\psi}$ defines the action for the spin sector.
Neglecting the quartic term $\mathcal{S}_4$ in Eq.~\eqref{sphi}, $\mathcal{S}_\varphi +
\mathcal{S}_{\varphi\psi}$ can be diagonalized numerically. The resulting spin
susceptibility $\chi''$ is finally averaged over typically 20 Monte-Carlo configurations.
While this procedure misses all inelastic scattering processes, it properly captures the
influence of short-range quasi-static stripe order on the magnetic modes. Clearly, it is
appropriate both for slowly fluctuating and for static but spatially disordered stripes
-- the latter is expected from impurity pinning which acts as a random field on the CDW
order parameter.

Although the theory $\mathcal{S}_{\varphi}+\mathcal{S}_{\psi}+\mathcal{S}_{\varphi\psi}$
has a number of input parameters, most can be fixed with an eye towards experiment. $c$
in $\mathcal{S}_{\varphi}$ sets the velocity of spin excitations and is known
experimentally; $s$ is then chosen as to reproduce the energy of the spin excitations at
$(\pi,\pi)$ (the ``spin resonance''). The overall scale of the parameters in
$\mathcal{S}_{\psi}$ sets the typical amplitude of the charge fluctuations $|\psi|_{\rm
typ}$. The product $\lambda|\psi|_{\rm typ}$ enters $\mathcal{S}_{\varphi\psi}$ and is
chosen such that the spin excitations become gapless at the incommensurate wavevector $\Qsp$
for perfect stripes. The interplay of the remaining parameters $c_{1,2;x,y}$,
$s_{x,y}$, $u_{1,2}$, $v$, and $w$ determines the character of the stripe fluctuations.
Here, reasonable combinations can be found be comparing the simulation result for
$\delta\rho = Q_x + Q_y$ with charge configurations deduced from STM data
\cite{kohsaka07}. The latter indicate that amplitude fluctuations are suppressed as
compared to phase fluctuations, that bond-centered pinning is strong, and that domain
walls are rather sharp. Finally, keeping all other parameters fixed, a variation of
$s_{x,y}$ will drive the charge sector through its order--disorder transition. We note
that the theory $\mathcal{S}_{\psi}$ \eqref{spsi} can be expected to have a nematic phase
at large positive $v$; however, we have not mapped out the full phase diagram of
$\mathcal{S}_{\psi}$.

Interestingly, Ref.~\cite{vvk} first determined plausible sets of parameters which were
able to reproduce an hour-glass spin fluctuation spectrum within the adiabatic
approximation to $\mathcal{S}_{\varphi}+\mathcal{S}_{\psi}+\mathcal{S}_{\varphi\psi}$
in a regime where the charge correlation length was between 10 and 50 lattice constants.
Later, the action $\mathcal{S}_{\psi}$, coupled to the single-particle sector of a
$d$-wave superconductor, was used to model electronic spectra of disordered-pinned
stripes. Here, the {\em same} parameters for $\mathcal{S}_{\psi}$ were used and good
agreement with the STM data of Ref.~\cite{kohsaka07} was obtained.

Spatial anisotropies arising from structural distortions of the CuO$_2$ layers enter the
theory at various places. Most important are the parameters of $\mathcal{S}_{\psi}$, in
particular if the charge sector is close to spontaneous rotational symmetry breaking and
hence sensitive to small anisotropies.
Below, we shall employ anisotropic mass terms, $s_x \neq s_y$, but anisotropic gradient
terms are feasible as well.

We close this section with a remark on Ref.~\cite{lawler10}, which proposed a mechanism
for incommensurate spin excitations induced by uniform nematic order. In fact, this
mechanism bears similarity to the one implemented in $\mathcal{S}_\varphi +
\mathcal{S}_{\varphi\psi}$ used here and in Ref.~\cite{vvk}: Using the notation of
Ref.~\cite{lawler10}, the gradient term in Eq.~(4) of Ref.~\cite{lawler10} has a negative
prefactor in one direction for $c_2^2 N > c_0^2$ -- the same happens in
$\mathcal{S}_\varphi + \mathcal{S}_{\varphi\psi}$ for large positive $\lambda_2$.
Microscopically, this translates into effectively ferromagnetic (instead of antiferromagnetic)
coupling, and higher-order gradients are required to render the theory well-defined.
However, from Ref.~\cite{lawler10} it remains unclear why uniform nematic order should
induce ferromagnetic effective couplings. Also, for the experimentally relevant
ordering wavevectors, the gradient expansion of Ref.~\cite{lawler10} is questionable, and
a lattice theory appears more appropriate.


\section{Spin excitations in the presence of lattice anisotropy}
\label{sec:res}

\begin{figure}[!t]
\begin{center}
\includegraphics[width=4.7in]{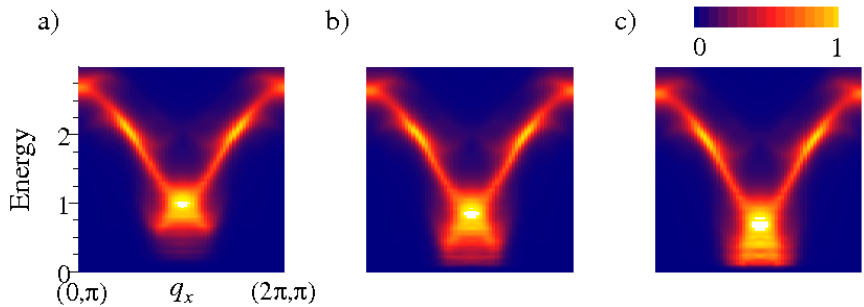}
\caption{
Neutron scattering intensity, $\chi''({\bf q},\omega)$ for momentum $(q_x,\pi)$, from Monte-Carlo
simulations of the coupled order parameter theory, Eq.~(\ref{sphi}--\ref{spsi}), for varying
spin-gap parameter $s$ in $\mathcal{S}_\varphi$ \cite{neg_foot,wgt_foot}.
The energy has been normalized to that of the ``resonance'' peak (which is roughly
35\,meV for the employed parameters \cite{param}) and an artificial energy broadening of
$\Gamma=0.05$ has been used.
Parameters are $s_x=s_y=-3.4$ and $u_1=-1.15$, placing $\mathcal{S}_\psi$ somewhat inside
the charge-ordered phase, and
a) $s=105$\,meV,
b) $s=88$\,meV,
c) $s=70$\,meV \cite{param}.
There is no in-plane anisotropy in the input parameters, and the spectra have been
symmetrized w.r.t. $q_x\leftrightarrow q_y$. The hour-glass shape is clearly visible.
}
\label{fig:scans}
\end{center}
\end{figure}

We now present numerical results for the magnetic excitation spectrum of disordered
stripes in the presence of lattice anisotropies. Although initial results were reported
in Ref.~\cite{vvk}, those were in a regime of a sizeable spin gap, and no attempt was
made to model the temperature evolution. In contrast, here we will choose parameters
with an eye towards the experimental data on YBCO-6.45 of Ref.~\cite{hinkov08a} where the
spin gap is small \cite{neg_foot}. Such variations of the spin gap are illustrated in
Fig.~\ref{fig:scans} where we show results for $\chi''(\vec q,\omega)$ with varying mass
parameter $s$ in $\mathcal{S}_\varphi$, here without in-plane anisotropy.

To quantify the incommensurability of the spin excitations, we shall analyze the
low-energy cuts through $\chi''(\vec q,\w)$ in the presence of an in-plane anisotropy in
a manner similar to Ref.~\cite{hinkov08a} in order to extract an apparent
incommensurability of the spin fluctuations. To assess the changes in the
spin-fluctuation spectrum with temperature, we shall vary the input parameters for
$S_\psi$ such that both amplitude and correlation length of stripe fluctuations change.

\begin{figure}[!t]
\begin{center}
\includegraphics[width=4.7in]{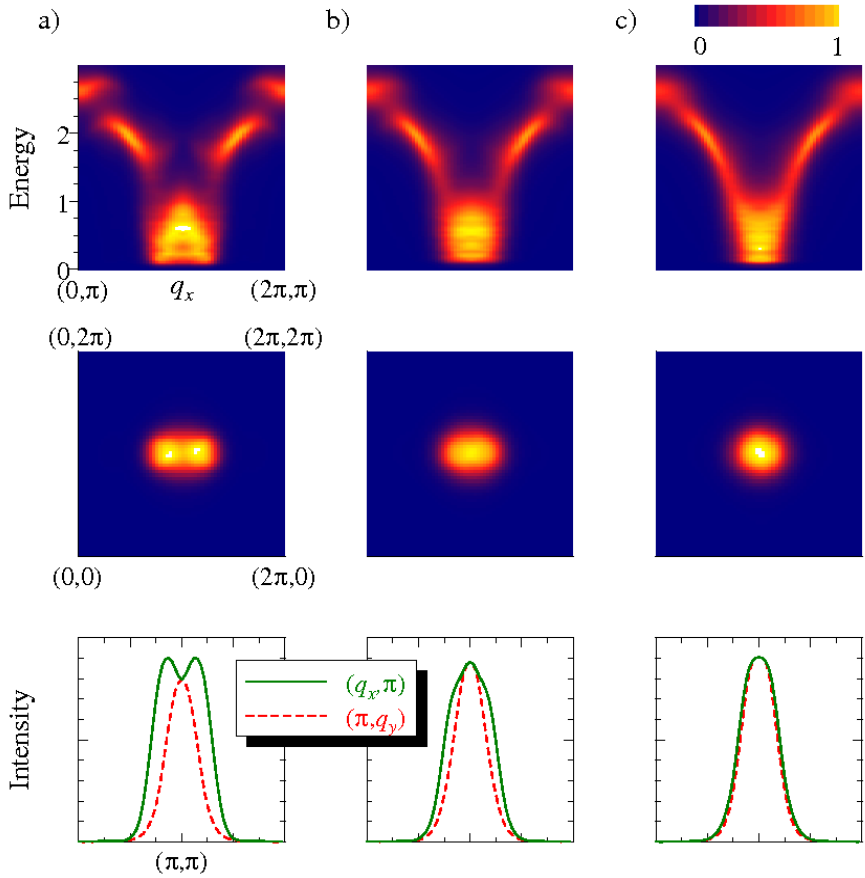}
\caption{
Results for $\chi''({\bf q},\omega)$ \cite{neg_foot,wgt_foot}
in the presence of an in-plane anisotropy for varying stripe correlation length.
Top: Color-coded intensity as function of energy and momentum $(q_x,\pi)$ along the $x$
axis.
Middle: Constant-energy scans at $\w=0.4$.
Bottom: Intensities along $(q_x,\pi)$ and $(\pi,q_y)$ at $\w=0.4$, showing the low-energy
anisotropy. The evolution from panels a to c mimics the increase of temperature: Here,
both $s_{x,y}$ and $u_1$ in the action $\mathcal{S}_{\psi}$ for the charge order
parameter are varied, such that the stripe correlation length $\xi$ decreases from a) to c).
The anisotropy is encoded in the mass difference for horizontal and vertical stripes,
$(s_x-s_y)$, and is held fixed, as is the spatial periodicity, $N=4$, of the underlying charge
stripes. Parameters are $s_y-s_x=0.01$,
a) $s_x=-2.8$, $u_1=-1.13$, $\xi\approx12$,
b) $s_x=-2.6$, $u_1=-1.11$, $\xi\approx9$,
c) $s_x=-2.0$, $u_1=-1.05$, $\xi\approx4$
\cite{param}.
}
\label{fig:chi1}
\end{center}
\end{figure}

The following assumptions on parameters will be made:
(a) The spatial period of the incipient charge order (and thus $\Qchx$, $\Qchy$) will be
held fixed --  this is motivated by the essentially temperature independent stripe
modulation period in 214 cuprates. For reasons of computational simplicity and momentum
resolution, we shall choose a charge period of 4 lattice constants. (Note that period 8
might be more appropriate for YBCO-6.45; as a result, our incommensurability of the spin
excitations will be roughly twice as large as in experiment.)
(b) We shall vary the strength of the charge order by varying $s_{x,y}$ and $u_1$ --
these parameters determine the correlation length and mean amplitude of $\psi_{x,y}$.
Qualitatively, these changes mimic the influence of temperature on the charge sector.
(Recall that the simulation of $\mathcal{S}_{\psi}$ is classical.)
Other parameters will be held fixed.
(c) The lattice anisotropy of YBCO will be implemented by anisotropic mass terms in  $\mathcal{S}_{\psi}$, $s_x
\neq s_y$. Near the ordering transition in the charge sector, this will induce strongly
anisotropic charge correlations, which in turn lead to anisotropies in the spin sector.
(Alternatively, anisotropic velocities in $\mathcal{S}_{\psi}$ can be employed.)
In principle, a small anisotropy should also be present in the bare spin action
$\mathcal{S}_{\varphi}$, but this will be neglected as the spin anisotropy is dominated
by the influence of the charge sector.

A central question is whether incipient charge order with a fixed spatial period can
induce incommensurate spin excitations with a temperature-dependent apparent
incommensurability. Ref.~\cite{lawler10} suggested that this is not the case (based on a
mean-field argument only), which then might rule out incipient stripe order as
explanation for the observation in Ref.~\cite{hinkov08a}.

A sequence of results for $\chi''(\vec q,\omega)$ with varying strength of the charge
order is shown in Fig.~\ref{fig:chi1}. For the parameters in panels b) and c), the
correlation length in the charge sector is 10 or smaller, and the coupling to the spin
sector produces a broad spin excitation spectrum at low energies, with a ``vertical
dispersion'', very similar to observations in underdoped YBCO \cite{hinkov08a,hinkov07}.
Note that moving closer to or inside the charge-ordered phase produces an hour-glass
spectrum in $\chi''(\vec q,\omega)$ (see Figs.~\ref{fig:scans} and \ref{fig:chi1}a as
well as Figs.~2,3 of Ref.~\cite{vvk}).

\begin{figure}[!t]
\begin{center}
\includegraphics[width=5.3in]{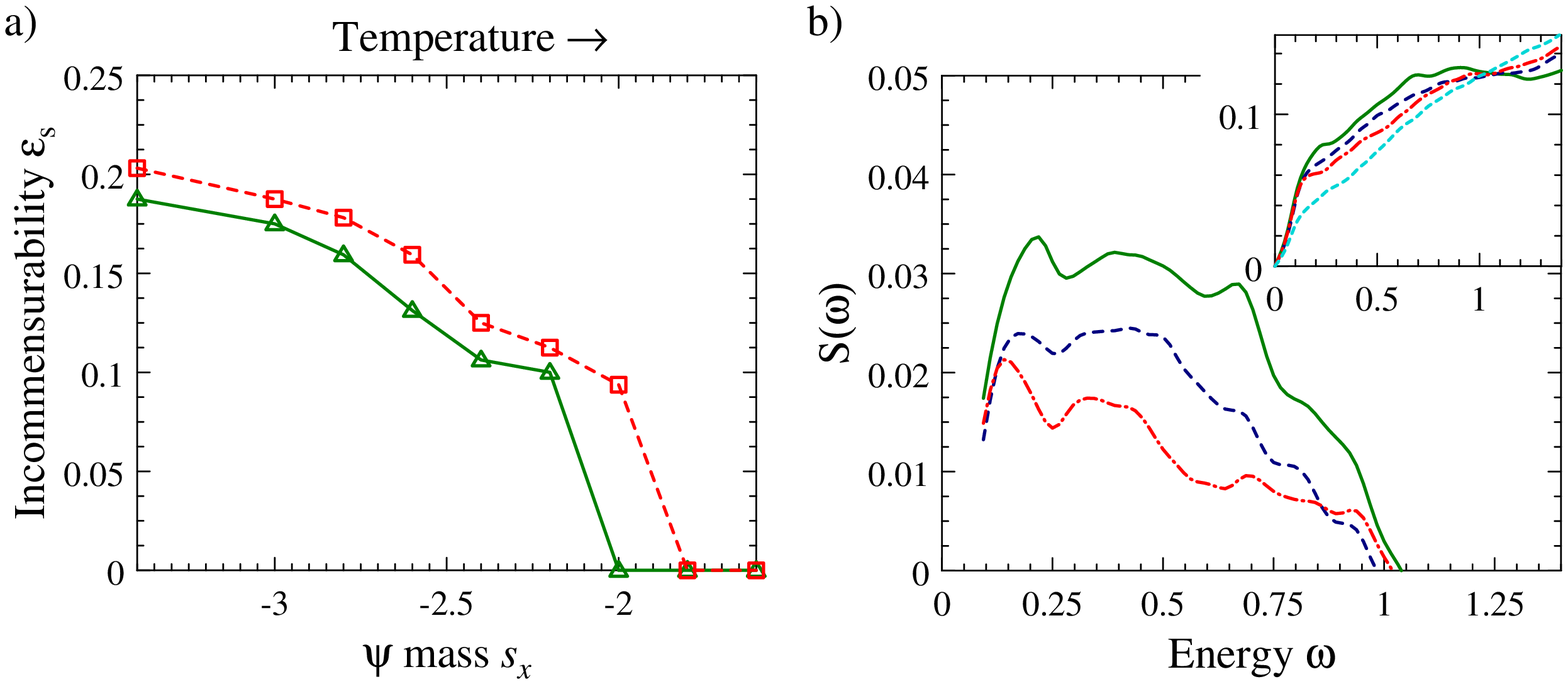}
\caption{
a) Apparent incommensurability $\epsilon_s$ of the low-energy intensity profile in $\chi''(\vec
q,\omega)$ as function of the order-parameter mass $s_x$; $u_1$ is varied along with
$s_x$ according to $u_1=s_x/10-0.85$. The two curves correspond to $s_y-s_x=0.01$ (solid)
and 0.02 (dashed); with the charge-ordering transitions located at $s_x$ values of
roughly 2.9 and 2.8, respectively.
$\epsilon_s$ is obtained from fitting $\chi''$ at $\w=0.4$ along $(q_x,\pi)$
to a sum of two Gaussians centered at $\pi(1\pm\epsilon_s,1)$.
($\epsilon_s$ is set to zero when the fit with a single Gaussian becomes visually
indistinguishable from the fit with two Gaussians.)
For perfect stripes, the incommensurability is $\epsilon_s=1/4$.
b) Momentum-integrated structure factor $S(\w)$ as function of energy for different stripe
correlation lengths. The inset shows the structure factor as obtained from the results
for $\chi''(\vec q,\omega)$, the main panel shows the same data with a ``background''
subtracted -- here the reference curve is chosen as the one with the shortest correlation
length in the inset. The background subtraction mimics the corresponding experimental
procedure, where the signal at high temperatures is typically smeared out due to a
combination of thermal fluctuations and fermionic damping and a linear background is
subtracted.
(The noise in all curves is due to Monte Carlo sampling.)
Parameters are $s_y-s_x=0.01$,
$s_x=-2.8$, $u_1=-1.13$ (solid),
$s_x=-2.4$, $u_1=-1.09$ (dashed),
$s_x=-2.0$, $u_1=-1.05$ (dash-dot),
$s_x=-1.0$, $u_1=-0.95$ (short dash)
\cite{param}.
}
\label{fig:sfac}
\end{center}
\end{figure}

The data in Fig.~\ref{fig:chi1} have been calculated in the presence of a small in-plane
anisotropy. Consequently, $\chi''(\vec q,\omega)$ at constant $\w$ below the
``resonance'' energy is anisotropic along the $x$ and $y$ directions, middle panel of
Fig.~\ref{fig:chi1}. The anisotropy at higher energies is less significant (not shown).
Following the experimental analysis of Ref.~\cite{hinkov08a}, we can fit the broad
intensity profile along the $x$ direction, bottom panel of Fig.~\ref{fig:chi1}, to a sum
of two Gaussians, centered at $\pi(1\pm\epsilon_s,1)$. The apparent incommensurability
$\epsilon_s$ resulting from such a fit is plotted in Fig.~\ref{fig:sfac}a as a function
of the mass of the charge order parameter. A number of features can be noted:
$\epsilon_s$ is finite both inside and outside the charge-ordered phase. It effectively
vanishes once the stripe correlation length becomes shorter than 5 lattice spacings (this
value depends on various parameters in $\mathcal{S}_\varphi +\mathcal{S}_{\varphi\psi}$),
and it approaches its ``saturation'' value $\epsilon_s=1/N$ dictated by the charge order
only deep in the ordered phase. Hence, even long-range ordered stripes do {\em not}
necessarily result in an apparent incommensurability of $\epsilon_s=1/N$.

Fig.~\ref{fig:sfac}a makes clear that the apparent incommensurability $\epsilon_s$ obtained from
fitting the profile of finite-energy spin fluctuations cannot be directly used to deduce
the spatial period of possibly underlying stripe order. In particular, a temperature
dependence of $\epsilon_s$ as observed in Ref.~\cite{hinkov08a} does not imply a
temperature dependence of the underlying stripe period, but can simply be a result of
smearing due to thermal and quantum fluctuations.

Finally, we show in Fig.~\ref{fig:sfac}b the momentum-integrated dynamic structure factor
$S(\w)$ obtained from the numerical simulations \cite{wgt_foot}. One observes a clear
transfer of spectral weight from high to low energies upon with increasing charge
correlation length, not unlike in the experimental data \cite{hinkov08a}. Note that, in
the present data, $S(\w)$ never displays a maximum at low energies (inset of
Fig.~\ref{fig:sfac}b); instead, a plateau emerges near the resonance energy which turns
into a peak for larger stripe correlation length. A low-energy peak in $S(\w)$ can only
be obtained after a ``background'' subtraction (main panel of Fig.~\ref{fig:sfac}b), here
a ``high-temperature'' curve has been subtracted. In the experiment, background
subtraction is done as well, with the difference that the high-temperature signal has
much less structure due to thermal spin fluctuations and fermionic damping, both of which
are absent from our simulation.


\section{Conclusions}

This paper has employed a scenario of slowly fluctuating or disorder-pinned stripes to
model the anisotropic spin excitation spectrum of underdoped YBCO. We have found that
such a theory, using plausible input parameters, is able to reproduce salient features of
the experiment of Ref.~\cite{hinkov08a}, including the temperature-dependent apparent
incommensurability in the low-energy spin excitations, in contrast to what has been
suggested in the literature.

Therefore, both stripe-driven and stripe-free nematic orders appear viable candidates to explain
the low-temperature anisotropies in YBCO. It remains to ask which experiments could
distinguish the two scenarios. Clearly, a direct observation of finite-$\vec q$ dynamical
charge fluctuations would support the stripe scenario, but a suitable experimental probe
does not exist. In the presence of quenched disorder, stripes will be pinned and should
be observable in real space, e.g., by STM. Unfortunately, clean surfaces as required for
STM are hard to obtain for this cuprate family. Resonant X-ray scattering can only detect
static charge order if the correlation length is sufficiently large, which is probably
not the case in YBCO \cite{stripe_rev2}. A more indirect probe is via the Fermi surface geometry. Whereas a
stripe-free nematic leads to a distorted single-sheet Fermi surface (at least in the
simplest tight-binding models), a stripe phase produces, depending on the periodicity, a
variety of smaller pockets \cite{millis_pockets,wollny08}.
Thus, photoemission experiments with sufficient energy
resolution might eventually be able to distinguish the two scenarios. (Note that for
fluctuating stripes those pockets will not be present at lowest energies; this is also of
relevance for the interpretation of Nernst data, where it has been argued \cite{hvs10} that {\em
static} stripes are incompatible with the large Nernst anisotropy seen at intermediate
temperatures \cite{taill10b}.)
Present-day quantum
oscillation data point toward the existence of pockets, albeit in a large magnetic field \cite{osc}.


\acknowledgement

I thank R. K. Kaul, S. Sachdev, T. Ulbricht, T. Vojta, and A. Wollny for collaborations
and fruitful discussions on this subject, E.-A. Kim, M. Lawler, and K. Sun for
correspondence regarding Ref.~\cite{lawler10}, and furthermore E. Fradkin, M. Greiter, V.
Hinkov, B. Keimer, S.~Kivelson, J.~Tranquada, and G.~Uhrig for discussions and
correspondence. This research was supported by the Deutsche Forschungsgemeinschaft
through the Research Unit FOR 538.



\begin{thebibliography}{}
\bibitem{neg_foot}
For small $s$ in ${\cal S}_{\varphi}$, charge disorder can lead to rare regions where
(locally) the spin gap becomes negative, reflecting the tendency to isolated islands with
weak magnetic order. For simplicity, we do not take into account this weak magnetic
condensate in our calculation, as it will influence the excitation spectrum only at the
very lowest energies. Hence, we simply ignore the few eigenmodes of $\mathcal{S}_\varphi
+ \mathcal{S}_{\varphi\psi}$ with negative eigenvalues.
\end{thebibliography}
\end{document}